\documentclass[a4paper,aps,onecolumn,nofootinbib]{revtex4}
\RequirePackage[english]{babel}
\RequirePackage[latin1]{inputenc}
\RequirePackage[T1]{fontenc}
\RequirePackage{mathrsfs}
\RequirePackage{amsmath}
\RequirePackage{amssymb}
\RequirePackage{amsbsy}
\RequirePackage{bm}
\usepackage{color}
\def\de#1/de#2{\frac{\partial {#1}}{\partial {#2}}}

\begin{document}
%%%%%%%%%%%%%%%%%%%%%%%%%%%%%%%%%%%%%%%%%%%%%%%%%%%%%%%%%%%%%%%%%%%%%%%%%%%%%%%%%%%%%%%%%%%%%%%%%%%
\title{\textbf{On the junction conditions in $f(R)$-gravity with torsion.}}
\author{Stefano Vignolo$^{1}$\footnote{E-mail: vignolo@dime.unige.it}, Roberto Cianci$^{1}$\footnote{E-mail: cianci@dime.unige.it}, Sante Carloni$^{2}$\footnote{E-mail: sante.carloni@tecnico.ulisboa.pt}}
\affiliation{$^{1}$DIME Sez. Metodi e Modelli Matematici, Universit\`{a} di Genova,\\
Piazzale Kennedy, Pad. D, 16129, Genova, ITALY\\
$^{2}$Centro de Astrof\'isica e Gravita\c{c}\~{a}o - CENTRA, Departamento de F\'isica,\\
Instituto Superior Técnico - IST, Universidade de Lisboa - UL,\\
Av. Rovisco Pais 1, 1049-001 Lisboa, Portugal.}
\date{\today}
%%%%%%%%%%%%%%%%%%%%%%%%%%%%%%%%%%%%%%%%%%%%%%%%%%%%%%%%%%%%%%%%%%%%%%%%%%%%%%%%%%%%%%%%%%%%%%%%%%%
\begin{abstract}
Junction conditions are discussed within the framework of $f(R)$-gravity with torsion. After deriving general junction conditions, the cases of coupling to a Dirac field and a spin fluid are explicitly dealt with. The main differences with respect to Einstein--Cartan--Sciama--Kibble theory $\left(f(R)=R\right)$ are outlined.   
\end{abstract}

%%%%%%%%%%%%%%%%%%%%%%%%%%%%%%%%%%%%%%%%%%%%%%%%%%%%%%%%%%%%%%%%%%%%%%%%%%%%%%%%%%%%%%%%%%%%%%%%%%%
\maketitle

%PACS numbers:

%%%%%%%%%%%%%%%%%%%%%%t%%%%%%%%%%%%%%%%%%%%%%%%%%%%%%%%%%%%%%%%%%%%%%%%%%%%%%%%%%%%%%%%%%%%%%%%%%%%%
\section{Introduction}\label{S1}
Higher order corrections to the Hilbert--Einstein action were recognized since long time as one of the main consequences of the introduction of quantum corrections to the matter sources of the gravitational field \cite{Birrell:1982ix}. Because of this feature, these additional terms were thought to be relevant only in the very early stages of the evolution of the universe and therefore were naturally linked to the inflationary paradigm. A model of this type that is still very much of interest is the so called Starobinsky model \cite{Starobinsky:1980te}. 

With the discovery of cosmic acceleration, the similarities between this phenomenon and the inflationary mechanism led the community to consider the possibility that dark energy could have a completely geometric origin \cite{GeomDE}. It was then recognized that the quantum higher order terms  have an effect on the entire cosmic history  and could be responsible of the measured increasing expansion rate of the universe at large scales. This idea was mainly explored at fourth order level and  more specifically in the various realizations of the so called $f(R)$-gravity theories \cite{f(R) review}.

If higher order corrections to the gravitational action indeed represent some approximation of the true (and yet undiscovered) quantum theory of gravitation, it should be natural that the matter sources of this theory possess the quintessential property of quantum matter: the spin. On the other hand,  the additional degrees of freedom that the spin introduces do not have any geometrical counterpart in classical Riemannian spacetimes. In the presence of spin, an extended geometrical framework is therefore needed. In order to provide such framework, Sciama  and Kibble \cite{Sciama1,Sciama:1964wt,Kibble:1961ba} suggested to abandon the Levi-Civita connection and to introduce torsion, thus providing the geometry of spacetime with the additional degrees of freedom to be coupled to  the quantum spin. The resulting theory, nowadays called Einstein-Cartan-Sciama-Kibble (ECSK) theory, has been extensively investigated since its initial formulation. 

Following the same paradigm of purely metric $f(R)$-gravity, ECSK theory has been generalized to $f(R)$-gravity with torsion \cite{Rubilar,CCSV1,CCSV2,Sotiriou,CV}. In such a theory, the main role played by the non--linearity of the gravitational Lagrangian function $f(R)$ is that to be a source for torsion, at the same time as the spin. 

So far, most of the applications of $f(R)$-gravity with torsion have been focused on cosmological models \cite{CCSV3,W,VFC,VF,CVF} and some of them have also concerned particle physics \cite{CDFV}. However, as in the majority of the extension of general relativity, standard cosmology does not provide constraints strong enough to evaluate completely the validity of a given theory. For this reason, the analysis of  the phenomenology of these theories in other sectors, like astrophysics, can yield to interesting new results. One common problem in this framework is connected with the matching of different spacetimes like, for example, in the case of the matching of the interior and the exterior of a relativistic stellar object. The requirements which need to be satisfied to solder two different spacetimes are commonly known as junction conditions.  

In standard general relativity  junction conditions were studied by Lichnerowicz \cite{L1,L2} and Israel \cite{Israel} (see also the works of Choquet--Bruhat \cite{Choquet} and Taub \cite{Taub}) and the solution of the problem is now well established (for a very clear discussion of the topic, the reader is referred to the book by Poisson \cite{Poisson} from which the authors borrowed arguments and notations). In the authors' knowledge, very few works were devoted to this topic in ECSK theory: an analysis was performed by Arkuszewski {\it et al.} \cite{AKP}, using the not immediate formalism of tensor--valued differential forms \cite{Trautman1,Trautman2}, while the same subject was indirectly addressed by Bressange \cite{Bressange} making use of the same approach as Poisson's. For what concerns $f(R)$-gravity, an analysis of junction conditions in the purely metric theory was proposed by Deruelle {\it et al.} \cite{Deruelle:2007pt} and Senovilla \cite{Senovilla}, while the topic has not yet been addressed within the theory with torsion.

To fill this gap, in this paper we generalize the formulation of the junction conditions to the case of $f(R)$-gravity with torsion. For the sake of brevity, in the following we will focus only on junction at spacelike or timelike separation surfaces: we will derive the junction conditions in general and we will discuss their explicit form in the case of coupling to a Dirac field and a spin fluid, which are two of the main matter fields with spin. As we shall see, the resulting junction conditions are only formally very similar to those of ECSK theory. Indeed, due to the non--linearity of the function $f(R)$, besides the spin the junction conditions also involve the trace of the stress--energy tensor and its first derivatives (evaluated on the separation hypersurface). This is a remarkable difference with respect to the ECSK theory that could have important consequences: for instance, in the case of coupling to a spin fluid, junction conditions could single out unacceptable state equations. 

The paper is organized as follows. In Sec. \ref{S2}, we recall the general features of $f(R)$-gravity with torsion. In Sec. \ref{S3}, we study the junction conditions in $f(R)$-gravity with torsion, outlining the differences with respect to ECSK theory. In Sec. \ref{S4}, we analyze in detail the cases of coupling to Dirac field and spin fluid. Finally, we devote Sec. \ref{S5} to conclusions. Throughout this paper, natural units ($\hbar=c=k_{B}=8\pi G=1$) are used and the signature $(+,-,-,-)$ is adopted.

\section{$f(R)$-gravity with torsion}\label{S2}
$f(R)$-gravity with torsion and spin has been formulated in terms of tetrad fields $e^\mu =e^\mu_i\,dx^i$ and spin connection $1$-forms $\omega^\mu_{\;\;\nu} = \omega_{i\;\;\;\nu}^{\;\;\mu}\,dx^i$ on the spacetime manifold $\cal M$ in \cite{CCSV2}. Greek letters denote Lorentz indexes which are raised and lowered by the Minkowski metric $\eta_{\mu\nu}={\rm diag.}\,(1,-1,-1,-1)$. The simultaneous assignment of a tetrad field and a spin--connection induces corresponding torsion and curvature tensors on $\cal M$, defined as
\begin{subequations}\label{2.1}
\begin{equation}\label{2.1a}
T^\mu = de^\mu + \omega^\mu_{\;\;\nu}\wedge e^\nu
\end{equation}
\begin{equation}\label{2.1b}
R^{\mu\nu} = d\omega^{\mu\nu} + \omega^\mu_{\;\;\lambda}\wedge\omega^{\lambda\nu}
\end{equation}
\end{subequations} 
In local coordinates, the components of torsion and curvature tensors are expressed as 
\begin{subequations}\label{2.2}
\begin{equation}\label{2.2a}
T^\mu_{ij} =\partial_i\/e^\mu_j - \partial_j\/e^\mu_i + \omega^{\;\;\mu}_{i\;\;\;\nu}e^\nu_{j} 
-\omega^{\;\;\mu}_{j\;\;\;\nu}e^\nu_{i}
\end{equation}
\begin{equation}\label{2.2b}
R_{ij}^{\;\;\;\;\mu\nu} =\partial_i\omega_{j}^{\;\;\mu\nu} - \partial_j{\omega_{i}^{\;\;\mu\nu}}
+\omega^{\;\;\mu}_{i\;\;\;\lambda}\omega_{j}^{\;\;\lambda\nu}
-\omega^{\;\;\mu}_{j\;\;\;\lambda}\omega_{i}^{\;\;\lambda\nu}
\end{equation}
\end{subequations}
Denoting by $R:= R_{ij}^{\;\;\;\;\mu\nu}\/e^i_\mu\/e^j_\nu$ (where $e^i_\mu$ denote the components of the dual of the tetrad $e^\mu$, satisfying $e^j_\mu\/e^\mu_i = \delta^j_i$ and $e^j_\mu\/e^\nu_j = \delta^\nu_\mu$) the curvature scalar written in terms of tetrad and spin connection, the field equations of $f(R)$-gravity with torsion are derived variationally through a Lagrangian density of the form
\begin{equation}\label{2.3}
{\cal L} = e\/f(R) - {\cal L}_m
\end{equation}
where $f(R)$ is a smooth function of the curvature scalar, ${\cal L}_m$ indicates a suitable matter Lagrangian coupled to gravity and $e:= \det \|e^\mu_i\|$. Variations with respect to tetrad and spin connection give rise to the field equations \cite{VC,CCSV2}
\begin{subequations}\label{2.4}
\begin{equation}\label{2.4a}
f'(R)\/R_{\mu\sigma}^{\;\;\;\;\lambda\sigma}e^{i}_\lambda -\frac{1}{2}\/f(R)\/e^i_{\mu}={\cal T}^i_\mu
\end{equation}
\begin{equation}\label{2.4b}
T_{ts}^{\;\;\;\alpha} = \frac{1}{2f'}\left(\de f'/de{x^p} + {\cal S}_{p\sigma}^{\;\;\;\sigma}\right)\left(\delta^{p}_{s}e^{\alpha}_{t}-\delta^{p}_{t}e^{\alpha}_{s}\right) + \frac{1}{f'}{\cal S}_{ts}^{\;\;\;\alpha}
\end{equation}
\end{subequations}
where ${\cal T}^{i}_{\;\mu}:=-\frac{1}{2e}\frac{\delta{\cal L}_m}{\delta e^{\mu}_{i}}$ and ${\cal S}_{ts}^{\;\;\;\alpha}:=\frac{1}{2e}\frac{\delta{\cal L}_m}{\delta\omega_{i}^{\;\;\mu\nu}}e^{\mu}_{t}e^{\nu}_{s}e^{\alpha}_{i}$ are the stress-energy and spin density tensors of the matter field, $R_{\mu\sigma}^{\;\;\;\;\lambda\sigma}e^{i}_\lambda := R_{ij}^{\;\;\;\;\lambda\sigma}\/e^i_\mu\/e^j_\sigma$, ${\cal S}_{p\sigma}^{\;\;\;\sigma}:={\cal S}_{pq}^{\;\;\;\sigma}e^q_\sigma$ and $f'$ denotes the first derivative of the function $f(R)$ with respect to $R$. In particular, eq. \eqref{2.4b} shows that the non--linearity of the function $f(R)$ is a source for the torsion together with the spin density of matter.  

By considering the trace of eqs. \eqref{2.4a}, we obtain a relation between the curvature scalar $R\/$ and the trace $\cal T\/$ of the stress--energy tensor given by
\begin{equation}\label{2.4bis}
f'\/(R)R -2f\/(R) = {\cal T}
\end{equation}
When the trace $\cal T\/$ is allowed to assume only a constant value, it is easily seen that the field equations of $f(R)$-gravity with torsion are formally equivalent to the ones of Einstein--Cartan theory with (or without) cosmological constant if spin is present, or the ones of Einstein theory with (or without) cosmological constant if spin is absent \footnote{When ${\cal T}$ is constant, the role of cosmological constant is played by the term $\frac{1}{\varphi(\cal T)}\/U(\cal T)$ introduced below and appearing in the subsequent eq. \eqref{2.7trisa}. This term can be zero depending on the adopted $f(R)$ model. For instance, it is zero for $f(R)=R+\alpha\/R^2$ in vacuo.}. This holds with the exception of the particular case ${\cal T}=0\/$ and $f(R)=\alpha\/R^2\/$. Indeed, under these conditions, eq. \eqref{2.4bis} is a trivial identity imposing no restrictions on the scalar curvature $R\/$.  

From now on, we shall systematically suppose that $\cal T\/$ is not forced to assume a constant value when the field equations are satisfied. Also, we shall suppose that the relation \eqref{2.4bis} is invertible so that the curvature scalar can be thought as a suitable function of $\cal T\/$, namely   
\begin{equation}\label{2.2.7}
R=R\/(\cal T)
\end{equation}

The implementation of $f(R)$ gravity with torsion presented in this paper, as well as in our previous works, relies on assumption \eqref{2.2.7}.
In this regard, it is worth noticing that, unlike what happens in the purely metric case, in metric--affine $f(R)$ theories of gravity the trace equation \eqref{2.4bis} defines an algebraic or transcendental relation between the curvature scalar and the stress--energy tensor trace, but not a differential one. Therefore, except for pathological situations, the Dini theorem is in general applicable and the relation \eqref{2.2.7} can be (almost always) supposed to locally exist. Among other things, this allows to express the torsion as a function of the matter fields and therefore to separate purely metric contributions from torsional ones within the Einstein-like equations, as it happens in the ECSK theory. 

Of course, the possibility of explicitly inverting eq. \eqref{2.4bis} and of obtaining analytically the expression \eqref{2.2.7} depends on the specific form of the function $f(R)$, but the point is that in general we are allowed to suppose to have the relation \eqref{2.2.7} at our disposal.  This turns out to be very useful for discussing general mathematical aspects of these theories such as the Cauchy problem or the junction conditions.

Metric compatibility conditions allow to associate to the spin connection a corresponding linear connection on the spacetime $\cal M$, given by
\begin{equation}\label{2.5}
\Gamma_{ij}^{\;\;\;h} = \omega_{i\;\;\;\nu}^{\;\;\mu}e_\mu^h\/e^\nu_j + e^{h}_{\mu}\partial_{i}e^{\mu}_{j}
\end{equation} 
Connection \eqref{2.5} defines the covariant derivative
\begin{equation}\label{2.6}
\nabla_{\partial_i}\partial_j = \Gamma_{ij}^{\;\;\;h}\,\frac{\partial }{\partial x^h}
\end{equation}
as well as the corresponding torsion and Riemann curvature tensors expressed as
\begin{subequations}\label{2.7}
\begin{equation}\label{2.7a}
T_{ij}^{\;\;\;h}
=\Gamma_{ij}^{\;\;\;h}-\Gamma_{ji}^{\;\;\;h}
\end{equation}
\begin{equation}\label{2.7b}
R^{h}_{\;\;kij}
=\partial_i\Gamma_{jk}^{\;\;\;h} - \partial_j\Gamma_{ik}^{\;\;\;h} +
\Gamma_{ip}^{\;\;\;h}\Gamma_{jk}^{\;\;\;p}-\Gamma_{jp}^{\;\;\;h}\Gamma_{ik}^{\;\;\;p}
\end{equation}
\end{subequations}
The relationships between the torsion and curvature components \eqref{2.7} and \eqref{2.2} are given by $T_{ij}^{\;\;\;h}=T^{\;\;\;\alpha}_{ij}e_\alpha^h$ and $R^{h}_{\;\;kij}=R_{ij\;\;\;\;\nu}^{\;\;\;\;\mu}e_\mu^h\/e^\nu_k$ respectively. Then, defining $R^i_{\;j}:=R_{\mu\sigma}^{\;\;\;\;\lambda\sigma}e^{i}_{\lambda}e^\mu_j$, ${\cal T}^i_{\;j}:={\cal T}^i_{\mu}e^\mu_j$ and ${\cal S}^{\;\;\;h}_{ij}:={\cal S}^{\;\;\;\alpha}_{ij}e_\alpha^h\/$, we can rewrite equations \eqref{2.4} in their equivalent general covariant form 
\begin{subequations}\label{2.7bis}
\begin{equation}\label{2.7bisa}
f'(R)\/R_{ij} -\frac{1}{2}f(R)\/g_{ij}
= {\cal T}_{ij}
\end{equation}
\begin{equation}\label{2.7bisb}
T_{ij}^{\;\;\;h} = \frac{1}{2f'}\left(\de f'/de{x^p} + {\cal S}_{pq}^{\;\;\;q}\right)\left(\delta^p_j\delta^h_i - \delta^p_i\delta^h_j\right) + \frac{1}{f'}{\cal S}^{\;\;\;h}_{ij}
\end{equation}
\end{subequations}
In equations \eqref{2.7bisa} one has to distinguish the order of the indexes, since in general the Ricci and energy--momentum tensors $R_{ij}=R^{h}_{\phantom{h}ihj}$ and ${\cal T}_{ij}\/$ are not symmetric.

Defining the scalar field $\varphi({\cal T}):= f'(R({\cal T}))$ and the quantity $U({\cal T}):=\left(\frac{1}{2}f(R({\cal T})) - \frac{1}{2}f'(R({\cal T}))\/R({\cal T})\right)$, eqs. \eqref{2.7bis} can be expressed in the equivalent form
\begin{subequations}\label{2.7tris}
\begin{equation}\label{2.7trisa}
R_{ij} - \frac{1}{2}R\/g_{ij}
= \frac{1}{\varphi}{\cal T}_{ij} + \frac{1}{\varphi}\/U\/g_{ij}
\end{equation}
\begin{equation}\label{2.7trisb}
T_{ij}^{\;\;\;h} = \frac{1}{2\varphi}\left(\de \varphi/de{x^p} + {\cal S}_{pq}^{\;\;\;q}\right)\left(\delta^p_j\delta^h_i - \delta^p_i\delta^h_j\right) + \frac{1}{\varphi}{\cal S}^{\;\;\;h}_{ij}
\end{equation}
\end{subequations}
which will be used in the following discussion.

Moreover, connection \eqref{2.5} is automatically metric compatible with respect to the metric tensor $g_{ij}=\eta_{\mu\nu}\/e^\mu_i\/e^\nu_j$ and can be decomposed as \cite{Hehl1,Hehl2}
\begin{equation}\label{2.8}
\Gamma_{ij}^{\;\;\;h}=\tilde{\Gamma}_{ij}^{\;\;\;h}-K_{ij}^{\;\;\;h}
\end{equation}
where 
\begin{equation}\label{2.9}
K_{ij}^{\;\;\;h}
:=\frac{1}{2}\left(-T_{ij}^{\;\;\;h}+T_{j\;\;\;i}^{\;\;h}-T^{h}_{\;\;ij}\right)
\end{equation}
is called the contorsion tensor and $\tilde{\Gamma}_{ij}^{\;\;\;h}$ is the Levi--Civita connection induced by the metric $g_{ij}$. Making use of eqs. \eqref{2.7trisb} and \eqref{2.9}, the contorsion tensor can be expressed in the form
\begin{subequations}\label{2.2.11}
\begin{equation}\label{2.2.11a}
K_{ij}^{\;\;\;h}= \hat{K}_{ij}^{\;\;\;h} + \hat{S}_{ij}^{\;\;\;h}
\end{equation}
\begin{equation}\label{2.2.11b}
\hat{S}_{ij}^{\;\;\;h}:=\frac{1}{2\varphi}\/\left( - {\cal S}_{ij}^{\;\;\;h} + {\cal S}_{j\;\;\;i}^{\;\;h} - {\cal S}^h_{\;\;ij}\right)
\end{equation}
\begin{equation}\label{2.2.11c}
\hat{K}_{ij}^{\;\;\;h} := -\hat{T}_j\delta^h_i + \hat{T}_pg^{ph}g_{ij}
\end{equation}
\begin{equation}\label{2.2.11d}
\hat{T}_j:=\frac{1}{2\varphi}\/\left( \de{\varphi}/de{x^j} + {\cal S}^{\;\;\;k}_{jk} \right)
\end{equation}
\end{subequations}
As a last remark, denoting by $T_i := T^{\;\;\;j}_{ij}$ the so--called torsion vector, we recall the conservation laws \cite{FV}
\begin{subequations}\label{2.10}
\begin{equation}\label{2.10a}
\nabla_{a}{\cal T}^{ai}+T_{a}{\cal T}^{ai}-{\cal T}_{ca}T^{ica}-\frac{1}{2}{\cal S}_{sta}R^{stai}=0
\end{equation}
\begin{equation}\label{2.10b}
\nabla_{h}{\cal S}^{ijh}+T_{h}{\cal S}^{ijh}+{\cal T}^{ij}-{\cal T}^{ji}=0
\end{equation}
\end{subequations}
under which the energy--momentum and spin density tensors of the matter fields must undergo. In particular, eqs. \eqref{2.10b} are equivalent to the antisymmetric part of the Einstein--like equations \eqref{2.7bisa}.

\section{The junction conditions}\label{S3}
Let $\Sigma$ be a hypersurface either timelike or spacelike \footnote{For brevity and simplicity, in this paper null hypersurfaces have been not considered. A further work will be devoted to the study of the junction conditions on null hypersurfaces in the framework of $f(R)$-gravity with torsion}
, separating two distinct regions ${\cal M}^+$ and ${\cal M}^-$ of spacetime. Let 
$\left(g_{ij}^+,\Gamma_{ij}^{+\;\;h}\right)$ and $\left(g_{ij}^-,\Gamma_{ij}^{-\;\;h}\right)$ be two different solutions of the field equations \eqref{2.7tris}.  We want to investigate the conditions under which the two given Einstein--Cartan geometries can be smoothly joined together at $\Sigma$, in such a way that $\left(g_{ij}^+,\Gamma_{ij}^{+\;\;h}\right)$ is solution of \eqref{2.7tris} in ${\cal M}^+$ and $\left(g_{ij}^-,\Gamma_{ij}^{-\;\;h}\right)$ in ${\cal M}^-$ respectively. Their matching forming then a solution of the field equations on the entire spacetime.

We suppose to cover $\Sigma$ with local coordinates $y^A$ ($A=1,\ldots,3$) and we imagine to work in a coordinate system $x^i$, locally overlapping both ${\cal M}^+$ and ${\cal M}^-$ in an open set containing $\Sigma$. For later use, we wish to define a function over spacetime, whose sign distinguishes ${\cal M}^+$ from ${\cal M}^-$. This can be achieved in different ways; for instance, for any point $p\in {\cal M}$ we can consider the arc length $s$ between $p$ and $\Sigma$ along the geodesic normal to $\Sigma$ (with respect to one of the two metrics) passing through $p$. Without loss of generality we can  set $s<0$ in ${\cal M}^-$, $s>0$ in ${\cal M}^+$ and $s=0$ at $\Sigma$. 

Denoting by $n^i$ the unit normal (with respect to the chosen metric) outgoing from $\Sigma$, we have the identities
\begin{equation}\label{3.0}
dx^i = n^i\/ds, \qquad n_i = \epsilon\partial_i\/s \qquad{\rm and}\qquad n^i\/n_i =\epsilon 
\end{equation}
with $\epsilon =1$ if $\Sigma$ is spacelike, $\epsilon =-1$ if $\Sigma$ is timelike. Moreover, given any geometric quantity $W$ defined on both sides of the hypersurface, the jump of $W$ across $\Sigma$ is denoted by
\begin{equation}
\left[W\right] := W\left({\cal M}^+\right)_{|\Sigma} - W\left({\cal M}^-\right)_{|\Sigma}
\end{equation}
We discuss the problem of matching solutions at $\Sigma$ in the framework of distribution--valued tensors \cite{L1,L2,Choquet,Dray1,Dray2}. To this aim, we introduce the Heaviside distribution $\Theta(s)$ (with $\Theta(0):=0$ \footnote{Other authors prefer to use the Heaviside distribution not defined in zero. Conclusions would be the same.}) and we 
define the geometrical objects
\begin{subequations}\label{3.1}
\begin{equation}\label{3.1a}
g_{ij} = \Theta(s)\/g_{ij}^+ + \left(1-\Theta(s)\right)\/g_{ij}^-
\end{equation}
\begin{equation}\label{3.1b}
\Gamma_{ij}^{\;\;\;h} = \Theta(s)\Gamma_{ij}^{+\;\;h} + \left(1-\Theta(s)\right)\Gamma_{ij}^{-\;\;h} 
\end{equation}
\end{subequations}
requiring  the quantities \eqref{3.1} to  represent an admissible solution of the Einstein--Cartan field equations \eqref{2.7tris} in distributional sense. It is therefore necessary that not only the quantities \eqref{3.1} are well defined as distributions, but also the geometric quantities connected to them such as the Riemann and Einstein tensors do. Moreover, consistency between \eqref{3.1a} and \eqref{3.1b} (see eq. \eqref{2.8}) requires that the connection \eqref{3.1b} has to coincide with the connection
\begin{equation}\label{3.2bb}
\Gamma_{ij}^{\;\;\;h} =  \Theta(s)\/\left(\tilde{\Gamma}_{ij}^{+\;\;\;h} -K_{ij}^{+\;\;h}\right) - \left[1-\Theta(s)\right]\/\left(\tilde{\Gamma}_{ij}^{-\;\;\;h} -K_{ij}^{-\;\;h}\right)
\end{equation}
where now $\tilde{\Gamma}_{ij}^{\;\;\;h}$ are the Christoffel coefficients associated with the metric \eqref{3.1a}.

That said, in view of the identities $\de{s}/de{x^i}= \epsilon\/n_i$ and $\frac{d\Theta(s)}{ds}=\delta(s)$ \footnote{For the definition of the Dirac $\delta$-function with support on the submanifold $\Sigma:s=0$, the reader is referred to \cite{Taub,Dray1,Dray2} and references therein.}, by differentiating \eqref{3.1} we have the relations
\begin{subequations}\label{3.2}
\begin{equation}\label{3.2a}
\partial_k\/g_{ij} = \Theta(s)\/\partial_k\/g_{ij}^+ + \left(1-\Theta(s)\right)\/\partial_k\/g_{ij}^- + \epsilon\/\delta(s)\left[g_{ij}\right]\/n_k
\end{equation}
\begin{equation}\label{3.2b}
\partial_k\/\Gamma_{ij}^{\;\;\;h} = \Theta(s)\partial_k\/\Gamma_{ij}^{+\;\;h} + \left(1-\Theta(s)\right)\partial_k\/\Gamma_{ij}^{-\;\;h} + \epsilon\delta(s)\/\left[\Gamma_{ij}^{\;\;\;h}\right]n_k 
\end{equation}
\end{subequations}
Taking the  identities $\Theta^2(s)=\Theta(s)$ and $\Theta(s)\left(1-\Theta(s)\right)=0$ into account,  it is easy to use the (\ref{3.2a}) to  calculate  the Levi-Civita contribution to the connection $\Gamma_{ij}^{\;\;\;h}$. This quantity will contain a singular term of the form
\begin{equation}\label{3.2bbb}
\frac{1}{2}g^{+hk}_{|\Sigma}\left(\left[g_{ik}\right]\/n_j + \left[g_{jk}\right]\/n_i - \left[g_{ij}\right]\/n_k\right)\/\epsilon\/\delta(s)
\end{equation}
Requiring that $\Gamma_{ij}^{\;\;\;h}$ has the form \eqref{3.1b}  implies the vanishing of the term \eqref{3.2bbb} and thus 
\begin{equation}\label{3.2bis}
\left[g_{ij}\right] =0
\end{equation} 
which implies the continuity of the two metrics across $\Sigma$. Next, using \eqref{3.2b} we can calculate the Riemann tensor associated with the connection \eqref{3.1b}
\begin{equation}\label{3.3}
R^{p}_{\;\;qij} = \Theta(s)R^{+p}_{\;\;\;\;qij} + \left(1-\Theta(s)\right)R^{-p}_{\;\;\;\;qij} + \delta(s)A^{p}_{\;\;qij} 
\end{equation}
where we have denoted by
\begin{equation}\label{3.4}
A^{p}_{\;\;qij} := \epsilon\left(\left[\Gamma_{jq}^{\;\;\;p}\right]\/n_i - \left[\Gamma_{iq}^{\;\;\;p}\right]\/n_j\right)
\end{equation}
the tensor responsible of the $\delta$-function term in the Riemann tensor \eqref{3.3}. As before, we can use the decomposition \eqref{2.8} so that the tensor \eqref{3.4} can be written as the sum
\begin{equation}\label{3.5}
A^{p}_{\;\;qij} = \tilde{A}^{p}_{\;\;qij} + \bar{A}^{p}_{\;\;qij}
\end{equation}
where
\begin{equation}\label{3.6}
\tilde{A}^{p}_{\;\;qij} = \epsilon\left(\left[\tilde{\Gamma}_{jq}^{\;\;\;p}\right]\/n_i - \left[\tilde{\Gamma}_{iq}^{\;\;\;p}\right]\/n_j\right)
\end{equation}
is the quantity due to the Levi--Civita connection, and
\begin{equation}\label{3.7}
\bar{A}^{p}_{\;\;qij} = \epsilon\left(\left[- K_{jq}^{\;\;\;p}\right]\/n_i + \left[K_{iq}^{\;\;\;p}\right]\/n_j\right)
\end{equation}
is the quantity due to contorsion. 

Now, since  the metric is continuous across the hypersurface $\Sigma$, any discontinuity of the derivatives of the metric may only exist along the normal direction. This means the existence of a tensor field on $\Sigma$
\begin{equation}\label{3.9}
k_{ij} := \epsilon\left[\partial_h\/g_{ij}\right]\/n^h 
\end{equation}
such that
\begin{equation}\label{3.8}
\left[\partial_h\/g_{ij}\right] = k_{ij}\/n_h
\end{equation}
Making use of \eqref{3.8}, we get the expressions
\begin{equation}\label{3.10}
\left[\tilde{\Gamma}_{ij}^{\;\;\;h}\right] = \frac{1}{2}\left(k^h_{\;\;j}\/n_i + k^h_{\;\;i}\/n_j - k_{ij}\/n^h\right)
\end{equation}
by which we obtain the explicit representation
\begin{equation}\label{3.11}
\tilde{A}^{p}_{\;\;qij} = \frac{\epsilon}{2}\left(k^p_{\;\;j}\/n_q\/n_i - k^p_{\;\;i}\/n_q\/n_j - k_{qj}\/n^p\/n_i + k_{qi}\/n^p\/n_j\right)
\end{equation}
Starting from eqs. \eqref{3.11}, by contraction we have
\begin{equation}\label{3.12}
\begin{split}
\tilde{A}_{qj} := \tilde{A}^{p}_{\;\;qpj}
=\frac{\epsilon}{2}\left(k^p_{\;\;j}\/n_q\/n_p - k\/n_q\/n_j - k_{qj}\epsilon + k_{qp}\/n^p\/n_j\right)
\end{split}
\end{equation}
and
\begin{equation}\label{3.13}
\tilde{A} := \tilde{A}^q_{\;\;q} = \epsilon\left(k_{pq}\/n^p\/n^q - \epsilon\/k\right)
\end{equation}
with $k:=k_{ij}\/g^{ij}$. Combining eqs. \eqref{3.12} with eqs. \eqref{3.13}, we define the tensor 
\begin{equation}\label{3.14}
\tilde{S}_{qj} = \tilde{A}_{qj} - \frac{1}{2}\tilde{A}g_{qj} = \frac{\epsilon}{2}\left(k^p_{\;\;j}\/n_q\/n_p - k\/n_q\/n_j - k_{qj}\epsilon + k_{qp}\/n^p\/n_j\right) - \frac{\epsilon}{2}\left(k_{st}\/n^s\/n^t - \epsilon\/k\right)\/g_{qj}
\end{equation}
representing the $\delta$-function part of the Einstein tensor, due to Levi--Civita contributions. By construction, tensor \eqref{3.14} is symmetric and tangent to the hypersurface $\Sigma$. Indeed, it is immediately seen that $\tilde{S}_{qj}\/n^j =0$. Then, denoting by $E^i_A :=\de{x^i}/de{y^A}$, $\tilde{S}_{qj}$ admits a representation of the form $\tilde{S}^{qj}=\tilde{S}^{AB}\/E^q_A\/E^j_B$, with \cite{Poisson}  
\begin{equation}\label{3.15}
\tilde{S}_{AB} := \tilde{S}_{qj}\/E^q_A\/E^j_B = -\frac{1}{2}k_{qj}\/E^q_A\/E^j_B + \frac{1}{2}k_{pq}h^{pq}\/h_{AB}
\end{equation}
where $h^{pq}:= g^{pq} -\epsilon\/n^p\/n^q$ is the projection operator on the hypersurface $\Sigma$ and $h_{AB}:=g_{ij}\/E^i_A\/E^j_B$ is the induced metric on $\Sigma$.

We can now evaluate the contributions given to the $\delta$-function part of the Einstein tensor by the contorsion terms. In fact, by contracting \eqref{3.7}, we have
\begin{equation}\label{3.16}
\bar{A}_{qj} := \bar{A}^{p}_{\;\;qpj} = \epsilon\left(\left[- K_{jq}^{\;\;\;p}\right]\/n_p + \left[K_{pq}^{\;\;\;p}\right]\/n_j\right)
\end{equation}
and
\begin{equation}\label{3.17}
\bar{A} := \bar{A}^q_{\;\;q} = 2\epsilon\left[K_{pq}^{\;\;\;p}\right]\/n^q
\end{equation}
On the basis of eqs. \eqref{3.16} and \eqref{3.17}, we define the tensor 
\begin{equation}\label{3.18} 
\bar{S}_{qj} := \bar{A}_{qj} - \frac{1}{2}\bar{A}g_{qj} = \epsilon\left(\left[- K_{jq}^{\;\;\;p}\right]\/n_p + \left[K_{pq}^{\;\;\;p}\right]\/n_j\right) - \epsilon\left[K_{st}^{\;\;\;s}\right]\/n^t\/g_{qj}
\end{equation}
It is worth noticing that in general tensor \eqref{3.18} is neither symmetric nor tangent to the hypersurface $\Sigma$. Collecting all the obtained results, we conclude that the effective stress--energy tensor on the right--hand side of eqs. \eqref{2.7trisa} can be written in the form
\begin{equation}\label{3.19}
\hat{{\cal T}}_{qj} = \Theta(s)\left[\frac{1}{\varphi}{\cal T}_{qj} + \frac{1}{\varphi}\/U\/g_{qj}\right]^+ 
 + \left(1-\Theta(s)\right)\left[\frac{1}{\varphi}{\cal T}_{qj} + \frac{1}{\varphi}\/U\/g_{qj}\right]^- - \delta(s)\/S_{qj}
\end{equation}
where
\begin{equation}\label{3.20}
S_{qj} = \tilde{S}_{qj} + \bar{S}_{qj}
\end{equation}
On the right--hand side of eq. \eqref{3.19}, the first and second terms are the energy--momentum tensors on the regions ${\cal M}^+$ and ${\cal M^-}$ respectively, while the $\delta$-function term can be interpreted as the energy--momentum tensor of a thin shell distribution of matter at $\Sigma$. Requiring a smooth transition across the hypersurface $\Sigma$ at the level of the Einstein--like equations amounts to imposing the vanishing at $\Sigma$ of the tensor $S_{qj}$ which is the source of the singular term in \eqref{3.19}. 

Therefore, in order to obtain the junction conditions, we need to study separately the vanishing of all orthogonal and tangent projections of $S_{qj}$ on $\Sigma$. In detail, we have: 
\begin{itemize}
\item since $\tilde{S}_{qj}$ is tangent to $\Sigma$ and the contorsion is antisymmetric in the last two indexes, the totally orthogonal projection of $S_{qj}$ on $\Sigma$ is automatically zero
\begin{equation}\label{3.21}
S_{qj}\/n^q\/n^j = \bar{S}_{qj}\/n^q\/n^j = - \epsilon\left[K_{j}^{\;\;qp}\right]\/n_p\/n_q\/n^j =0
\end{equation}
\item the tangent--orthogonal projection of $S_{qj}$ is
\begin{equation}\label{3.22}
S_{qj}\/E^q_A\/n^j = \bar{S}_{qj}\/E_A^q\/n^j = -\epsilon\left[K_{jq}^{\;\;\;p}\right]\/n_p\/E^q_A\/n^j + \left[K_{pq}^{\;\;\;p}\right]\/E^q_A
\end{equation}
According to \cite{AKP}, the quantity in eq. \eqref{3.22} is the jump of the projection on $\Sigma$ (or the pull-back on $\Sigma$) of the trace of the contorsion tensor. Indeed, we have the identity 
\begin{equation}\label{3.22bis}
\left[K_{jq}^{\;\;\;p}\/h^j_i\/h^i_p\/E^q_A\right] = \left[K_{jq}^{\;\;\;p}\/\left(\delta^j_p -\epsilon\/n_p\/n^j\right)\/E^q_A\right] = -\epsilon\left[K_{jq}^{\;\;\;p}\right]\/n_p\/n^j\/E^q_A + \left[K_{pq}^{\;\;\;p}\right]\/E^q_A
\end{equation} 
\item again due to the antisymmetry properties of the contorsion tensor and the orthogonality between the vectors $n^i$ and $E^i_A$, the orthogonal--tangent projection of $S_{qj}$ is zero 
\begin{equation}\label{3.23}
S_{qj}\/E^j_A\/n^q = \bar{S}_{qj}\/E_A^j\/n^q = \epsilon\left(-\left[K_{jq}^{\;\;\;p}\right]\/n_p\/n^q\/E^j_A + \left[K_{pq}^{\;\;\;p}\right]n_j\/E^j_A\/n^q\right) = 0
\end{equation}
\item finally, the totally tangent projection of $S_{qj}$ is given by
\begin{equation}\label{3.24}
S_{qj}\/E^q_A\/E^j_B = \tilde{S}_{qj}\/E^q_A\/E^j_B + \bar{S}_{qj}\/E^q_A\/E^j_B = \tilde{S}_{AB} + \epsilon\left(- \left[K_{jq}^{\;\;\;p}\right]\/n_p\/E^q_A\/E^j_B + \left[K_{q}^{\;\;qp}\right]\/n_p\/h_{AB}\right)
\end{equation}
\end{itemize}
Summarizing, we see that the vanishing of the tensor $S_{qj}$ requires that the quantities \eqref{3.22} and \eqref{3.24} are zero at $\Sigma$. In particular, as  in the purely metric case (General Relativity), the condition $S_{qj}\/E^q_A\/E^j_B =0$ can be related  to the vanishing of the jump of the extrinsic curvature across $\Sigma$. In order to see this point, we introduce the quantity
\begin{equation}\label{3.25}
Q_{AB} := \left(\nabla_i\/n_j\right)\/E^j_A\/E^i_B 
\end{equation}
which generalizes the notion of extrinsic curvature for an arbitrary linear connection, including the ones like \eqref{2.8}. By eq. \eqref{3.25} together with eqs. \eqref{2.8} and \eqref{3.10}, we deduce the relation
\begin{equation}\label{3.26}
\left[Q_{AB}\right] = \left[\nabla_i\/n_j\/E^j_A\/E^i_B \right] = \left[\nabla_i\/n_j\right]\/E^j_A\/E^i_B = \frac{\epsilon}{2}\/k_{ij}\/E^i_A\/E^j_B + \left[K_{ji}^{\;\;\;h}\right]\/n_h\/E^i_A\/E^j_B
\end{equation}
Now, making use of eq. \eqref{3.26}, it is a straightforward matter to verify the identity 
\begin{equation}\label{3.27}
S_{AB} := S_{kj}\/E^k_A\/E^j_B = - \epsilon\left(\left[Q_{AB}\right] - \left[Q\right]\/h_{AB}\right) 
\end{equation}
where $\left[Q\right]:= \left[Q_{AB}\right]\/h^{AB}$. 

Hence, in complete analogy with what is happening in Einstein theory \cite{Poisson}, we conclude that the requirement $S_{AB}=0$ at $\Sigma$ is equivalent to the vanishing at $\Sigma$ of the jump $\left[Q_{AB}\right]$ of extrinsic curvature, this time associated with the dynamical connection \eqref{2.8} instead of the Levi--Civita one. 

As a final remark, it is worth noticing that the vanishing of the quantities \eqref{3.22} and \eqref{3.24} involves not only the Levi--Civita connection and the spin tensor (through the contorsion tensor) as it happens in ECSK theory, but also the trace of the energy--impulse tensor and its first derivatives. This is due to the contributions that the function $f(R)\not = R$ gives to torsion. This aspect is clarified by the two examples proposed in the next section.

\section{Coupling to Dirac field and spin fluid}\label{S4}
Two main kinds of matter fields with spin, usually coupled to gravity, are the Dirac field and the spin fluid. In this Section, we examine in detail the conditions arising from the vanishing of expressions \eqref{3.22} and \eqref{3.24} in the case of coupling to these two matter fields.

\subsection{Spin fluid}
In the case of coupling to a Weyssenhoff spin fluid, the stress--energy and the spin tensors are respectively given by \cite{Hehl2,Obukhov,VF} 
\begin{subequations}\label{3.30.1}
\begin{equation}\label{3.30.1a}
{\cal T}^{ij}=U^i\/P^j +p\left(U^i\/U^j - g^{ij}\right)
\end{equation}
and
\begin{equation}
{\cal S}_{ij}^{\;\;\;h} = {\cal S}_{ij}\/U^h
\end{equation}
\end{subequations}
where $U^i$ and $P^j$ denote respectively the $4$-velocity and  the $4$-density of energy--momentum, ${\cal S}_{ij}=-{\cal S}_{ji}$ is the spin density and $p$ is the pressure of the fluid. Making use of the conservation law for the spin \eqref{2.10b}, which amounts to the validity of the antisymmetric part of Einstein--like equations \eqref{2.7trisa}, the stress--energy tensor \eqref{3.30.1a} can be expressed in the form \cite{VF}
\begin{equation}
{\cal T}_{ij}= \left(\rho + p\right)\/U_i\/U_j -p\/g_{ij} - U_i\/{\hat T}_h\/{\cal S}^h_{\;\;j} - U_i\tilde{\nabla}_h\left({\cal S}_{kj}U^h\right)\/U^k
\end{equation}
where $\rho:=U^iP_i$ and $\tilde{\nabla}_h$ denotes covariant derivative with respect to the Levi--Civita connection induced by the metric $g_{ij}$. Taking the usual convective condition \cite{Obukhov,Prasanna} ${\cal S}_{ij}\/U^j =0$ into account and making use again of the representation \eqref{2.2.11}, a direct calculation shows that the vanishing at $\Sigma$ of the quantities \eqref{3.22} and \eqref{3.24} gives rise to the explicit equations
\begin{subequations}\label{3.31}
\begin{equation}\label{3.31a}
-\epsilon\left[K_{jq}^{\;\;\;p}\right]\/n_p\/E^q_A\/n^j + \left[K_{pq}^{\;\;\;p}\right]\/E^q_A = -\epsilon\left[\frac{1}{\varphi}{\cal S}_{qj}\/U_p\right]\/n^p\/n^j\/E^q_A - \left[\frac{1}{\varphi}\de\varphi/de{x^q}\right]\/E^q_A =0
\end{equation}
\begin{equation}\label{3.31b}
\begin{split}
\tilde{S}_{AB} + \epsilon\left(- \left[K_{jq}^{\;\;\;p}\right]\/n_p\/E^q_A\/E^j_B + \left[K_{q}^{\;\;qp}\right]\/n_p\/h_{AB}\right) = \\
\tilde{S}_{AB} + \epsilon\left(\left[\frac{1}{2\varphi}\left({\cal S}_{jq}U^{p} + {\cal S}^p_{\;\;q}U_j + {\cal S}^{p}_{\;\;j}U_q\right)\right]\/n_p\/E^q_A\/E^j_B + \left[\frac{1}{\varphi}\de\varphi/de{x^p}\right]\/n^p\/h_{AB}\right) =0
\end{split}
\end{equation}
\end{subequations}
Separating the symmetric part from the antisymmetric one of \eqref{3.31b}, we finally obtain the conditions
\begin{subequations}\label{3.32}
\begin{equation}\label{3.32a}
\left[\frac{1}{2\varphi}{\cal S}_{jq}U^{p}\right]\/n_p\/E^q_A\/E^j_B =0
\end{equation}
\begin{equation}\label{3.32b}
\tilde{S}_{AB} + \epsilon\left(\left[\frac{1}{2\varphi}\left({\cal S}^p_{\;\;q}U_j + {\cal S}^{p}_{\;\;j}U_q\right)\right]\/n_p\/E^q_A\/E^j_B + \left[\frac{1}{\varphi}\de\varphi/de{x^p}\right]\/n^p\/h_{AB}\right) =0
\end{equation}
\end{subequations}
Let us now apply these results in a simple specific case. Let us consider the junction of two static and spherically symmetric metrics 
\begin{equation}\label{3.28}
ds_{\pm}^2 = e^{\nu^{\pm}}\,dt^2 - e^{\lambda^{\pm}}\,dr^2 - r^2\left(d\theta^2 + \sin^2\theta\,d\phi^2\right)
\end{equation}
which are solutions of eqs. \eqref{2.7tris} coupled to a spin fluid. In this case, renaming the spherical coordinates as $x^0:=t,x^1:=r,x^2:=\theta,x^3:=\phi$ for simplicity, the $4$-velocity of the fluid (at rest in the chosen frame) is described by $U^i = U^0\/\delta^i_0$, with $U^0 = e^{-\frac{\nu}{2}}$, and the unit normal to the hypersurface $\Sigma : x^1={\rm const.}$ is given by $n^i = n^1\/\delta^i_1$ with $n^1 = e^{-\frac{\lambda}{2}}$. The functions $\nu$ and $\lambda$, as well as all the involved matter fields, depend only on the radial variale $r$. 

The convective condition ${\cal S}_{ij}\/U^j =0$ and the assumed spherical symmetry imply that the only nonzero components of the spin density are ${\cal S}_{23}=-{\cal S}_{32}$. This condition is equivalent  to the  requirement that the spins of the particles composing the fluid are all aligned in the $r$ direction \cite{Prasanna}. Moreover, in this case the stress--energy tensor of the spin fluid assumes the usual form of the stress--energy tensor of a perfect fluid 
\begin{equation}\label{3.33.1}
{\cal T}_{ij} = \left(\rho + p\right)\/U_i\/U_j - p\/g_{ij}  
\end{equation} 
Using the above assumptions, the constraints \eqref{3.31a} and \eqref{3.32a} are automatically satisfied, while eq. \eqref{3.32b} reduces to
\begin{equation}\label{3.33}
\tilde{S}_{AB} + \epsilon\left[\frac{1}{\varphi}\de\varphi/de{x^p}\right]\/n^ph_{AB} =0
\end{equation} 
which relates the quantity $\left[\frac{1}{\varphi}\de\varphi/de{x^p}\right]$ to the jump across $\Sigma$ of the extrinsic curvature $\tilde{Q}_{AB}$ associated with the Levi--Civita connection of the metric \eqref{3.28}. We have in fact the identity \cite{Poisson} $\tilde{S}_{AB} = -\epsilon\left(\left[\tilde{Q}_{AB}\right] - \left[\tilde{Q}\right]\/h_{AB}\right)$. We notice that, in the case of Einstein--Cartan theory ($\varphi=1$), condition \eqref{3.33} becomes $\tilde{S}_{AB}=0$ (amounting to $\left[\tilde{Q}_{AB}\right]=0$) which coincides with the condition holding in General Relativity \cite{Poisson}.

Because of \eqref{2.4bis} and the definition of $\varphi$, we expect that condition \eqref{3.33} involves the derivatives of matter fields (or, at least, some of them). To clarify this point, we consider for instance the model $f(R)= R+\alpha\/R^2$, where $\alpha$ is a suitable constant. In view of eq. \eqref{3.33.1}, the trace equation \eqref{2.4bis} yields the relation
\begin{equation}\label{3.34}
- R ={\cal T} =\rho -3p
\end{equation}
which implies
\begin{equation}\label{3.35}
\varphi = 1+2\alpha\left(3p-\rho\right)
\end{equation}
In addition to this, it is an easy matter to see that the only non--vanishing component of the extrinsic curvature $\tilde{Q}_{AB}$ (induced by the metric \eqref{3.28}) of the hypersurface $\Sigma: \quad r=r_0 \quad {\rm const.}$, is given by
\begin{equation}\label{3.36}
\tilde{Q}_{00} = {\frac{1}{2}\de\nu/de{r}e^{\nu-\frac{\lambda}{2}}}_{|r=r_0}
\end{equation}
It follows that requirement \eqref{3.33} amounts to the two distinct conditions
\begin{subequations}\label{3.37}
\begin{equation}\label{3.37a}
\left[\frac{2\alpha\left(3\de p/de{r}-\de \rho/de{r}\right)}{1+2\alpha\left(3p-\rho\right)}\right] =0
\end{equation}
and
\begin{equation}\label{3.37b}
\left[\de\nu/de{r}\right]=0
\end{equation}
\end{subequations} 
As a specific example, we assume that the region ${\cal M}^+$ of spacetime is empty. Typically, we can think of joining together the interior with the exterior spacetime of a star with spin properties.
In such a circumstance $({\cal T}_{ij}^+=0, {\cal S}_{ij}^{+\;h}=0)$, in ${\cal M}^+$ the field equations \eqref{2.7tris} are identical to the Einstein equations (without cosmological constant) in vacuo and therefore they admit as unique solution $\left(g_{ij}^+,\Gamma_{ij}^{+\,\;h}\right)$, the Schwartzchild metric 
\begin{equation}\label{3.38}
g_{ij}^+\,dx^idx^j = \left(1-\frac{2M}{r}\right)\,dt^2 - \left(1-\frac{2M}{r}\right)^{-1}\,dr^2 - r^2\left(d\theta^2 + \sin^2\theta\,d\phi^2\right)
\end{equation} 
together with its Levi--Civita connection $\Gamma_{ij}^{+\;\;h}=\tilde{\Gamma}_{ij}^{+\;\,h}$. In view of this, the junction conditions \eqref{3.2bis} and \eqref{3.37} become
\begin{subequations}\label{3.39}
\begin{equation}\label{3.39a}
e^{\nu^{-}(r_0)} = \left(1-\frac{2M}{r_0}\right), \qquad e^{\lambda^{-}(r_0)} = \left(1-\frac{2M}{r_0}\right)^{-1}, 
\end{equation}
\begin{equation}\label{3.39b}
\left(\de\nu^{-}/de{r}\right)_{|r=r_0}={\frac{2M}{r_0\left(r_0 -2M\right)}}, \qquad {\left(3\de p^{-}/de{r}-\de \rho^{-}/de{r}\right)}_{|r=r_0} =0
\end{equation}
\end{subequations}
The first condition can be used to determine the mass constant of the exterior Schwarzschild solution. The \eqref{3.39b} can be also used, via the field equations, to translate  junction conditions in constraints on the thermodynamics of the fluid on the junction hypersurface. It is clear that, since the pressure term contains derivatives of the metric function which are not regulated by \eqref{3.39b}, the pressure is not automatically zero as it happens in GR. On the other hand, if the junction border we consider is the hypersurface of a compact object the condition $p=0$ is simply required by the request of stability of the object. This means that, in the present framework, for compact objects the condition $p=0$ on the separation hypersurface has to be imposed by a suitable choice of the constants appearing in the interior solution. This is very different from the standard GR case and could be used as a way to test these theories on compact objects.
Finally the second of \eqref{3.39b} implies that the spin fluid must have the  barotropic factor $w=\left(\de p^{-}/de{\rho^{-}}\right)_{|r=r_0}=1/3$ at the boundary, i.e. it has to behave like a sort of radiation at $\Sigma$. 

The results we obtained are, in fact, quite general. In particular, for all $f(R)$ models with torsion admitting static and spherically symmetric solutions \eqref{3.28}, the condition ${\left(-3\de p^{-}/de{r} + \de\rho^{-}/de{r}\right)}_{|r=r_0} ={\de{{\cal T}^{-}}/de{r}}_{|r=r_0}=0$ is always sufficient (together with \eqref{3.37b}) to fulfill the requirement \eqref{3.33}, and it becomes also necessary whenever ${\frac{\partial\varphi^-}{\partial{\cal T}}}_{|r=r_0}\not =0$ (like in the case $f(R)=R+\alpha\/R^2$, where ${\frac{\partial\varphi^-}{\partial{\cal T}}}_{|r=r_0}=-2\alpha$). On the other hand, whenever the condition ${\frac{\partial\varphi^-}{\partial{\cal T}}}_{|r=r_0}=0$ is imposed, it automatically generates a relation between density and pressure at the separation hypersurface, constraining the admissible equations of state. The same reasoning in principle can be used in any theory in which an algebraic-transcendental coupling between $R$ and ${\cal T}$ is present, i.e. in principle for any of such theories the junction conditions might constrain the equation of state of the fluid. This could happen, for example, for $f(R)$-gravity in the Palatini formulation.

%%%%%%%%%%%%%%%%%%%%%%%%%%%%%%%%%%%%%%%%%%%%%%%%%%%%%%%%%%%%%%%%%%%%%%%%%%
%%%%%%%%%%%%%%%%%%%%%%%%%%%%%%%%%%%%%%%%%%%%%%%%%%%%%%%%%%%%%%%%%%%%%%%%%%

\subsection{Dirac field}
We now consider the case of $f(R)$-gravity with torsion coupled to a Dirac field $\psi$. The matter Lagrangian function is the Dirac one
\begin{equation}\label{DL}
{\cal L}_m=
\left[\frac{i}{2}\left(\bar\psi\gamma^iD_{i}\psi-D_{i}\bar\psi\gamma^{i}\psi\right)-m\bar\psi\psi\right]
\end{equation}
where $D_i\psi = \de\psi/de{x^i} + \omega_i^{\;\;\mu\nu}S_{\mu\nu}\psi\/$ and $D_i\bar\psi = \de{\bar\psi}/de{x^i} - \bar\psi\omega_i^{\;\;\mu\nu}S_{\mu\nu}\/$ are the covariant derivatives of the Dirac fields, $S_{\mu\nu}=\frac{1}{8}\left[\gamma_\mu,\gamma_\nu\right]\/$, $\gamma^i =\gamma^{\mu}e^i_\mu\/$ with $\gamma^\mu\/$ denoting Dirac matrices and where $m$ is the mass of the Dirac field. The notation for which 
\begin{equation}\label{notation}
\gamma^{\mu}\gamma^{\nu}\gamma^{\lambda}
=\gamma^{\mu}\eta^{\nu\lambda}-\gamma^{\nu}\eta^{\mu\lambda}+\gamma^{\lambda}\eta^{\mu\nu}
+i\epsilon^{\mu\nu\lambda\tau}\gamma_{5}\gamma_\tau
\end{equation}
is used. The Dirac equations derived from \eqref{DL} are
\begin{equation}\label{DE}
i\gamma^{h}D_{h}\psi + \frac{i}{2}T_h\gamma^h\psi- m\psi=0
\end{equation}
Note the presence in eq. \eqref{DE} of the contribution given by the torsion vector $T_h:=T_{hj}^{\;\;\,j}$. This is due to the non--linearity of the gravitational Lagrangian function $f(R)$, so to the fact that torsion is no longer proportional to spin (thus no longer totally antisymmetric). Making use of \eqref{DE}, it easily seen that the stress--energy tensor is given by
\begin{equation}\label{SET}
{\cal T}_{ij}=\frac{i}{4}\left(\bar\psi\gamma_{i}D_{j}\psi-D_{j}\bar\psi\gamma_{i}\psi\right)
\end{equation}
while the totally antisymmetric spin density tensor  can be expressed as \cite{Hehl2,FV}
\begin{equation}\label{3.27bisb}
{\cal S}_{ij}^{\;\;\;h}= -\frac{1}{4}\eta^{\mu\sigma}\epsilon_{\sigma\nu\lambda\tau}
\left(\bar{\psi}\gamma_{5}\gamma^{\tau}\psi\right)e^{h}_{\mu}e^{\nu}_{i}e^{\lambda}_{j}
\end{equation}
In view of representation \eqref{2.2.11}, it is an easy matter to see that in the case of coupling to Dirac field the vanishing at $\Sigma$ of the quantities \eqref{3.22} and \eqref{3.24} reduces to the conditions
\begin{subequations}\label{3.29}
\begin{equation}\label{3.29a}
-\epsilon\left[K_{jq}^{\;\;\;p}\right]\/n_p\/E^q_A\/n^j + \left[K_{pq}^{\;\;\;p}\right]\/E^q_A = -\epsilon\left[\hat{K}_{jq}^{\;\;\;p}\right]\/n_p\/E^q_A\/n^j + \left[\hat{K}_{pq}^{\;\;\;p}\right]\/E^q_A = - \left[\frac{1}{\varphi}\de\varphi/de{x^q}\right]\/E^q_A =0
\end{equation}
\begin{equation}\label{3.29b}
\begin{split}
\tilde{S}_{AB} + \epsilon\left(- \left[K_{jq}^{\;\;\;p}\right]\/n_p\/E^q_A\/E^j_B + \left[K_{q}^{\;\;qp}\right]\/n_p\/h_{AB}\right) = \\
\tilde{S}_{AB} + \epsilon\left(- \left[\hat{K}_{jq}^{\;\;\;p}\right]\/n_p\/E^q_A\/E^j_B + \left[\hat{K}_{q}^{\;\;qp}\right]\/n_p\/h_{AB} - \left[\hat{S}_{jq}^{\;\;\;p}\right]\/n_p\/E^q_A\/E^j_B\right) = \\
\tilde{S}_{AB} + \epsilon\left(\left[\frac{1}{\varphi}{\cal S}_{jq}^{\;\;\;p}\right]\/n_p\/E^q_A\/E^j_B + \left[\frac{1}{\varphi}\de\varphi/de{x^p}\right]\/n^p\/h_{AB}\right) =0
\end{split}
\end{equation}
\end{subequations}
Eq. \eqref{3.29b} can be splitted in its symmetric and antisymmetric parts, so obtaining the equations
\begin{subequations}\label{3.30}
\begin{equation}\label{3.30a}
\tilde{S}_{AB} + \epsilon\left[\frac{1}{\varphi}\de\varphi/de{x^p}\right]\/n^p\/h_{AB} =0
\end{equation}
\begin{equation}\label{3.30b}
\left[\frac{1}{\varphi}{\cal S}_{jq}^{\;\;\;p}\right]\/n_p\/E^q_A\/E^j_B =0
\end{equation}
\end{subequations}
As an explicit example, we consider again the model $f(R)= R+\alpha\/R^2$ and suppose to join two axially symmetric spacetimes, solutions of the resulting field equations. The choice of axial symmetry instead of spherical symmetry is due to the fact that in gravity with torsion there can not be non-trivial spinors filling a spherically symmetric spacetime \cite{Luca}. 

Therefore, in both the regions ${\cal M}^-$  and ${\cal M}^+$, we assume the corresponding metrics to be of Lewis--Papapetrou kind in spherical coordinates
\begin{equation}\label{Lewis}
g^\pm_{ij}\,dx^i\/dx^j = -B_\pm^2\/(r^2\,d\theta^2 +dr^2) -A_\pm^2\/(-W_\pm\,dt + d\phi)^2 + C_\pm^2\,dt^2
\end{equation}
where all functions $A_\pm(r,\theta)$, $B_\pm(r,\theta)$, $C_\pm(r,\theta)$ and $W_\pm(r,\theta)$ depend on the $r$ and $\theta$ variables only. We suppose that the region ${\cal M}^+$ is empty while the region ${\cal M^-}$ is filled with a Dirac field.  Some examples of axially symmetric solutions of the form \eqref{Lewis} in gravity coupled to Weyl fermions can be found in a work by some of us \cite{CFV}, but here their explicit form is not important and will not be discussed in detail. On the other hand, in the region ${\cal M}^+$ we choose the Kerr metric. This is justified by the fact that $R+\alpha\/R^2$ gravity with torsion in vacuo amounts to Einstein gravity and therefore admits the same solutions. In the Lewis--Papapetrou form \eqref{Lewis}, the Kerr metric reads
\begin{subequations}\label{ABCW}
\begin{equation}\label{A}
A^2_+(r,\theta) = \left[a^2 + \frac{\left(-a^2 + m^2 +2mr + r^2\right)^2}{4r^2}\right]\sin^2\theta + \frac{ma^2\left(-a^2 + m^2 +2mr + r^2\right)\sin^4\theta}{r\left(\frac{\left(-a^2 + m^2 +2mr + r^2\right)^2}{4r^2} + a^2\cos^2\theta\right)}
\end{equation}
\begin{equation}\label{B}
B^2_+(r,\theta) = \frac{a^2\cos^2\theta}{r^2} + \frac{1}{4} + \frac{m}{r} + \frac{3m^2-a^2}{2r^2} + \frac{m^3-a^2m}{r^3} + \frac{a^4 -2a^2m^2 + m^4}{4r^4}
\end{equation}
\begin{equation}\label{C}
\begin{split}
C^2_+(r,\theta) = \frac{m^2a^2\left(-a^2 + m^2 +2mr + r^2\right)^2\sin^4\theta}{\left(\left(a^2 + \frac{\left(-a^2 + m^2 +2mr + r^2\right)^2}{4r^2}\right)\sin^2\theta + 
\frac{ma^2\left(-a^2 + m^2 +2mr + r^2\right)\sin^4\theta}{r\left(\frac{\left(-a^2 + m^2 +2mr + r^2\right)^2}{4r^2} + a^2\cos^2\theta\right)}\right)}\times\\
\frac{1}{r^2\left(\frac{\left(-a^2 + m^2 +2mr + r^2\right)^2}{4r^2} + a^2\cos^2\theta\right)^2} +1 - \frac{m\left(-a^2 + m^2 +2mr + r^2\right)}{r\left(\frac{\left(-a^2 + m^2 +2mr + r^2\right)^2}{4r^2} + a^2\cos^2\theta\right)}
\end{split}
\end{equation}
\begin{equation}\label{W}
\begin{split}
W_+(r,\theta) = \frac{ma\left(-a^2 + m^2 +2mr + r^2\right)\sin^2\theta}{\left(a^2+\frac{\left(-a^2 + m^2 +2mr + r^2\right)^2}{4r^2}\right)\sin^2\theta + \frac{ma^2\left(-a^2 + m^2 +2mr + r^2\right)\sin^4\theta}{r\left(\frac{\left(-a^2 + m^2 +2mr + r^2\right)^2}{4r^2}+a^2\cos^2\theta\right)}}\times\\
\frac{1}{r\left(\frac{\left(-a^2 + m^2 +2mr + r^2\right)^2}{4r^2}+a^2\cos^2\theta\right)}
\end{split} 
\end{equation}
\end{subequations}
where $a$ and $m$ are the parameters entering the Kerr metric. We want to discuss now the junction conditions at the hypersurface $\Sigma: \quad r=r_0 \quad {\rm const}$. To this end, by using the Dirac equations \eqref{DE} to evaluate the trace of the stress--energy tensor \eqref{SET}, we preliminary observe that in the region ${\cal M^-}$ we have the identity
\begin{subequations}\label{varphi}
\begin{equation}\label{varphi-}
\varphi^- = 1+ 2\alpha\/R = 1- 2\alpha{\cal T} = 1- \alpha\/m\bar\psi\psi
\end{equation}
while in the region ${\cal M^+}$ we have 
\begin{equation}\label{varphi+}
\varphi^+ =1
\end{equation}
\end{subequations}
Taking eqs. \eqref{varphi} into account, it is evident that the constraint \eqref{3.29a} implies that the scalar $\bar\psi\psi$ has to be constant on the hypersurface $\Sigma$. After that, a direct calculation shows that the requirement \eqref{3.30b} amounts to the further conditions
\begin{equation}\label{spinvector}
{\bar\psi\gamma_5\gamma^0\psi}_{|\Sigma}=0, \qquad  {\bar\psi\gamma_5\gamma^2\psi}_{|\Sigma}=0, \qquad  {\bar\psi\gamma_5\gamma^3\psi}_{|\Sigma}=0
\end{equation}
that the spinor field $\psi$ has to verify at $\Sigma$. Finally, in order to discuss the remaining condition \eqref{3.30a}, we first rewrite it in the equivalent form
\begin{equation}\label{equazionebis}
\left[\tilde{Q}_{AB}\right] = - \frac{1}{2}\left[\frac{1}{\varphi}\frac{\partial\varphi}{\partial x^h}\right]\/n^h\/h_{AB}
\end{equation}
where $\left[\tilde{Q}_{AB}\right]$ denotes the jump across $\Sigma$ of the extrinsic curvatures induced by the metrics \eqref{Lewis}. Now, denoting by $\tilde{A}:=A^+(r_0,\theta)=A^-(r_0,\theta)$, $\tilde{B}:=B^+(r_0,\theta)=B^-(r_0,\theta)$, $\tilde{C}:=C^+(r_0,\theta)=C^-(r_0,\theta)$ and $\tilde{W}:=W^+(r_0,\theta)= W^-(r_0,\theta)$ for simplicity, the non--vanishing components of $\left[\tilde{Q}_{AB}\right]$ result to be the following ones
\begin{subequations}\label{K}
\begin{equation}\label{K22}
\left[\tilde{Q}_{\theta\theta}\right] = -r_0^2\left[\partial_r\/B\right]
\end{equation}
\begin{equation}\label{K33}
\left[\tilde{Q}_{\phi\phi}\right] = -\frac{\tilde{A}}{\tilde{B}}\left[\partial_r\/A\right]
\end{equation}
\begin{equation}\label{K34}
\left[\tilde{Q}_{t\phi}\right] = \frac{\tilde{A}\left(2\tilde{W}\left[\partial_r\/A\right] + \tilde{A}\left[\partial_r\/W\right]\right)}{2\tilde{B}}
\end{equation}
\begin{equation}\label{K44}
\left[\tilde{Q}_{tt}\right] = \frac{\tilde{C}\left[\partial_r\/C\right] - \tilde{A}\tilde{W}^2\left[\partial_r\/A\right] - \tilde{A}^2\tilde{W}\left[\partial_r\/W\right]}{\tilde{B}}
\end{equation}
\end{subequations}
In view of eqs. \eqref{varphi} and \eqref{K}, the non--trivial equations of \eqref{equazionebis} assume the explicit form
\begin{subequations}\label{equazionetris}
\begin{equation}\label{equazionetrisa}
\frac{\left[\partial_r\/B\right]}{\tilde{B}} = - \frac{\alpha\/m}{2(1-\alpha\/m\bar\psi\psi_{|\Sigma})}\partial_r\left(\bar\psi\psi\right)_{|\Sigma}
\end{equation}
\begin{equation}\label{equazionetrisb}
\frac{\left[\partial_r\/A\right]}{\tilde{A}} = - \frac{\alpha\/m}{2(1-\alpha\/m\bar\psi\psi_{|\Sigma})}\partial_r\left(\bar\psi\psi\right)_{|\Sigma}
\end{equation}
\begin{equation}\label{equazionetrisc}
\frac{2\left[\partial_r\/A\right]}{\tilde{A}}  + \frac{\left[\partial_r\/W\right]}{\tilde{W}}= - \frac{\alpha\/m}{(1-\alpha\/m\bar\psi\psi_{|\Sigma})}\partial_r\left(\bar\psi\psi\right)_{|\Sigma}
\end{equation}
\begin{equation}\label{equazionetrisd}
\frac{\tilde{C}\left[\partial_r\/C\right] - \tilde{A}\tilde{W}^2\left[\partial_r\/A\right] - \tilde{A}^2\tilde{W}\left[\partial_r\/W\right]}{\tilde{C}^2 - \tilde{A}^2\tilde{W}^2} = - \frac{\alpha\/m}{2(1-\alpha\/m\bar\psi\psi_{|\Sigma})}\partial_r\left(\bar\psi\psi\right)_{|\Sigma}
\end{equation}
\end{subequations}
In particular, from eqs. \eqref{equazionetris} it follows that the jumps of the $r$-derivatives of quantities $A^\pm$, $B^\pm$, and $C^\pm$ have to satisfy the relations
\begin{equation}\label{jumps}
\frac{\left[\partial_r\/A\right]}{\tilde{A}} = \frac{\left[\partial_r\/B\right]}{\tilde{B}} = \frac{\left[\partial_r\/C\right]}{\tilde{C}} =- \frac{\alpha\/m}{2(1-\alpha\/m\bar\psi\psi_{|\Sigma})}\partial_r\left(\bar\psi\psi\right)_{|\Sigma} = \frac{1}{2\varphi}\de\varphi/de{(\bar\psi\psi)}\partial_r\left(\bar\psi\psi\right)_{|\Sigma}
\end{equation}
while the function $W(r,\theta)$ have to be of class ${\cal C}^1$. 

Summarizing the obtained results, we see that in the case of ECSK theory ($f(R)=R+\lambda$, thus $\varphi =1$) the metric resulting from the matching has to be at least of class ${\cal C}^1$; on the contrary, in the general case ($f(R)\not= R+\lambda$) the derivatives of the metric components with respect to the coordinate $r$ can have some jumps which are related to the $r$--derivative of the scalar quantity $\bar\psi\psi$ of the Dirac field. Moreover, it is shown that, in addition to the requirements \eqref{spinvector} imposed on the spin vector as it would occur in the ordinary ECSK theory, the junction conditions in $f(R)$-gravity with torsion also involve the scalar field $\bar\psi\psi$ through equations \eqref{3.29a} and \eqref{jumps}. Therefore, also the behavior of the scalar $\bar\psi\psi$ at the separation hypersurface $\Sigma$ contributes to the determination of the junction constraints.

\section{Conclusions}\label{S5}
In this paper we have derived the junction conditions of two generic spacetimes in the context of $f(R)$-gravity with torsion. In particular, we have shown that these conditions can be recast into a form that resemble very closely that holding in the ECSK theory \cite{AKP,Bressange}, via the definition of a suitable ``effective extrinsic curvature'' tensor. Using such a tensor one can both deduce the conditions of junction and determine the stress energy tensor $S_{kj}$ of the shell corresponding to the violation of such conditions. 

The close similarity to the case of ECSK theory, however, is only formal. Indeed, due to the contributions that the non--linearity of the gravitational Lagrangian function $f(R)$ gives to the contorsion tensor, in the obtained junction conditions there is the presence of the trace of the stress--energy tensor and its first derivatives. In general, all this results in suitable constraints imposed on the matter fields at the separation hypersurface and represents a remarkable difference with respect to the ECSK theory. In order to acquire a more profound understanding of this aspect, we have considered two specific examples assuming the model $f(R)=R+\alpha R^2$.  

In the first example, we have considered as source a Weyssenhoff fluid. In this case, the junction conditions imposed on the scalar field $\varphi$ imply very specific requirements on the energy density and pressure of the spin fluid. In the case of a spherically symmetric interior metric soldered to the Schwarzschild solution, i.e. the case of a compact object, one obtains two interesting results. First, differently from GR, the junction conditions do not lead necessarily to zero pressure at the separation surface. Since the stability of the compact object depends strictly on the fact that the pressure is zero on the surface of separation, this implies that a consistent object requires the fulfillment of an additional constraint. In addition, the conditions on the derivatives of the energy density and the pressure coming from the junction constrain the barotropic factor of the fluid at the border. In particular, and rather unexpectedly, we obtained that such a constraint is $w=1/3$ which represents a form of radiation. In other words, at the border the spin fluid has to behave as form of radiation to be able to perform the junction. What kind of matter is compatible with this result?  Certainly the fluid must be relativistic, but at the same time retaining a non negligible spin interaction, as the fluid remains Weyssenhoff no matter the speed of sound. Such a situation can be only physically meaningful if the fluid is very dense at the surface. On the other hand, one could redo the calculations using a standard perfect fluid and obtain the same results. In such a circumstance, the requirement of high density would be unnecessary. We should also stress that we have assumed that the source matter is ``cold'' in the sense that the properties of the fluid do not depend on the temperature. In a more realistic case (that goes well beyond the purpose of this example), the properties of the fluid could change with the temperature/speed of sound leading to a weakening of these constraints.

In the second example, instead, a classical massive Dirac field represents the matter source of the field equations. Since it is known that in the presence of torsion no spherically symmetric solutions of the gravitational field equations coupled to a non--trivial Dirac field can exist \cite{CFV}, we considered the case of a cylindrical interior solution of the  Lewis--Papapetrou kind soldered to an exterior Kerr metric. In this case the junction conditions are seen to constrain the behavior of the spinor field at the separation hypersurface. In addition to that, the junction of the metric coefficients results to be related to the values of the scalar field $\bar\psi\psi$ and its radial first derivative at the hypersurface of separation. 

All in all, our results point to the fact that the request of junction of two given solutions in the context of $f(R)$-gravity with torsion requires some very specific conditions on the behavior of the matter fields on the separation hypersurface. These differences with respect to ECSK theory are like to be translated to features that can be actually observed. Of course, the details of these conditions will depend on the features of the solutions that need to be soldered. Future works will be dedicated to the search of these solutions and their detailed analysis.

\begin{acknowledgments}
S. C.~was supported by  the Funda\c{c}\~{a}o para a Ci\^{e}ncia e Tecnologia through project IF/00250/2013 and partly funded through H2020 ERC Consolidator Grant ``Matter and strong-field gravity: New frontiers in Einstein's theory'', grant agreement no. MaGRaTh-64659.

The Authors would like to thank anonymous Referees for their useful comments contributing to improve the paper.
\end{acknowledgments}
%%%%%%%%%%%%%%%%%%%%%%%%%%%%%%%%%%%%%%%%%%%%%%%%%%%%%%%%%%%%%%%%%%%%%%%%%%%%%%%%%%%%%%%%%%%%%%%%%%%%%%%%%%%%%

\end{document}